\documentclass{article}

\usepackage{amssymb,amsmath,mathtools,xcolor,graphicx,xspace,colortbl,ragged2e,rotating} %
\usepackage{amsmath}  
\usepackage{boxedminipage}  
\usepackage{float}  
\usepackage{ragged2e}  
\usepackage{xcolor}  
\graphicspath{{retrocausality_castagnoli july 5 2025_graphics/}{retrocausality_castagnoli july 5 2025_tcache/}{retrocausality_castagnoli july 5 2025_gcache/}}
\DeclareGraphicsExtensions{.pdf,.eps,.ps,.png,.jpg,.jpeg}

\begin{document}
\title{Quantum computational speedup and retrocausality
}
\author{G. Castagnoli\\$$Formerly: Elsag Bailey ICT Division and Quantum Laboratory\protect\footnote{
giuseppe.castagnoli@gmail.com
}
}
\maketitle
\begin{abstract}Involving only the measurements of commuting observables -- the problem-setting and the corresponding solution -- quantum algorithms should be subject to classical logic. This would allow flanking their customary quantum description with a classical logic description, with surprising consequences. In the classical  logic description of the quantum algorithm, very simply, it is as if the problem-solver knew in advance, before beginning her problem-solving action, one of the possible halves of the information that specifies the solution of the problem she will produce and measure in the future and could use this knowledge to produce the solution with fewer computation steps. This is a causal loop whose retrocausal character turns out to be implicit in the very notion of quantum state superposition, both an essential ingredient of the quantum computational speedup and one of the pillars of quantum mechanics. Indeed, the key point of the work is that the classical logic description of a quantum state superposition must resort to a logical form of retrocausality that in turn must be physically implicit in the superposition itself. The existence of retrocausality in ordinary quantum physics implies e different way of viewing physical reality. It explains in a unified way all quantum speedups and quantum nonlocality. It highlights the teleological character of quantum algorithms, that is, their being evolutions toward a goal (the solution of the problem) with an attractor in the solution they will produce in the future (the solution again). Under the quantum cosmological assumption, it provides a plausible physical basis to the teleological character of natural evolutions.
\end{abstract}

\section{Concept
}
The present work is little mathematics and mostly a different way of viewing the quantum computational speedup. Before going on to formalizations, we thought it appropriate to provide a conceptual exposition of the view.

\subsection{Introduction
}
In the evolutionary approach of ours and co-authors $\left [1 -5\right ]$, we formulated the assumption that the quantum computational speedup is explained by a logical causal loop. In it, it is as if the problem-solver knew in advance, before beginning her problem-solving action, one (anyone) of the possible halves of the information about the solution she will produce and measure in the future and could use this knowledge to produce the solution with fewer computation steps. In the optimal quantum case, the number of computation steps would be that required to produce the solution in an optimal logical way benefiting of the advanced knowledge in question.

More precisely, the quantum speedup is explained by the mutually exclusive $or$ of the logical causal loops in question, each for each possible way of taking half of the information about the solution.

We called this ``the advanced knowledge rule'', and showed that it fits all the major quantum algorithms $\left [4\right ]$. Note that it also establishes a very direct, quantitative, comparison between quantum and classical computation: in the optimal case, the number of computation steps needed by the quantum algorithm is that of a classical algorithm that benefits of the advanced knowledge of one of the possible halves of the information about the solution of the problem.

We also showed that the customary quantum description of quantum algorithms, which says nothing of the causal loops in question, would consequently be incomplete and would be completed by time-symmetrizing it. This leaves the unitary part of the customary quantum description (the computation of the solution) mathematically unaltered but changes the description of the behavior of causality along it. The single causal process of the customary quantum description is replaced by the quantum superposition of the causal loops whose mutually exclusive $or$ explains the quantum computational speedup. In this way, the logical causal loops we are dealing with become physical, quantum indeed.

By the way, the need to time-symmetrize the ordinary quantum description of quantum algorithms will also be confirmed by a symmetry implicit in this description. This goes in favor of the assumption that quantum algorithms are subject to classical logic and guarantees its consequence.

Until now, the advanced knowledge rule we are dealing with has been an empirical fact satisfied by all the quantum algorithms examined. In this work, we should have theoretically demonstrated it, incidentally in a surprisingly simple way. By the way, concerning retrocausality, it would be a demonstration with broad consequences on our way of viewing the physics of both the microscopic and the macroscopic phenomena.

We can start from the observation that, only involving the measurements of commuting observables -- the problem-setting and the corresponding solution of the problem -- quantum algorithms should be subject to classical logic. Therefore, their ordinary quantum description could be flanked by a classical logic description and, of course, the two descriptions should go along. As we will soon see, the classical logic description of what the quantum algorithm does -- its producing a certain quantum speedup -- very simply implies the advanced knowledge rule we are dealing with. Indeed, it immediately says that it is as if the the problem solver knew in advance one of the possible halves of the information about the solution she will produce and measure in the future and could use this knowledge to produce the solution with a quantum speedup.

This is, of course, a causal loop, which in turn implies retrocausality. Naturally, it must be a retrocausality implicit -- and hidden -- in the ordinary quantum description of the quantum algorithm and made explicit by its classical logic description. 

Now a quantum algorithm is a rather complex physical phenomenon. A natural question is then whether there is a more elementary quantum phenomenon whose classical logic description implies retrocausality. Of course, this is indeed the case. The phenomenon in question is quantum state superposition, both an essential ingredient of the quantum speedup and one of the pillars of quantum mechanics.

Let us show how its classical logic description must resort to retrocausality. This is shown in a most simple way by resorting to a prescientific notion of state superposition -- it will be the same with the ordinary quantum description.

It is the ancient legend of the monk seen by his disciples on the two banks of a river at the same time. Now, from a logical standpoint, these two possibilities, of being on one or the other bank, are mutually exclusive with one another. Apparently, being on both banks at the same time is thus a logical impossibility. However, instead, it becomes possible if one opens to retrocausality, as follows. The monk, first, follows a path that leads him to one of the two banks of the river; then (``then'' in causal order) he goes sufficiently backward in time undoing the path in question (as in a reversed film), then he goes forward in time again but following a different path that brings him to the other bank. With this causality zigzag, he can, of course, have been seen on the two banks of the river at the same time. We can also see that any classical logic description of this state superposition inevitably involves retrocausality. This is the rather simple key of the entire work.

We will see how the above classical logic description of a state superposition can be ported, as it is, into the quantum world where it explains the quantum speedup in a unified quantitative way for all oracle quantum computing, and much more.

\subsection{
Classical logic description of Grover algorithm
}
Before going to show that the classical logic description of an optimal quantum speedup implies the advanced knowledge rule we are dealing with, let us make a few remarks on the latter.

This rule provides a unified quantitative explanation of an important class of quantum speedups. Given an oracle problem (any one), it provides the number of computation steps (oracles queries) needed to solve it in an optimal quantum way; it is the number needed to solve it in an optimal logical way benefiting of the advanced knowledge of one of the possible halves of the information about the solution.

In today's quantum information, the only way of knowing the number of oracle queries needed to solve an oracle problem in an optimal quantum way is finding the quantum algorithms that does that. Of course, this is an object of high ingenuity, uncertain achievement, and different from oracle problem to oracle problem. As far as we know, excluding extremely simple quantum algorithms with just one computation step, this object has been achieved only once, indeed in the case of the optimal quantum search algorithm of Grover $\left [6\right ]$.

Despite its potential convenience, the rule in question obviously cannot fit into today's quantum information. As quantum mechanics itself, the latter relies on the fundamental notion of causal evolution, namely an evolution where causality always goes in the same time-direction\protect\footnote{
We will be dealing with time-reversible evolutions, where causality can either go forward or backward in time. However, in causal evolutions, the two directions are mutually exclusive with one another.
}. By the way, all forms of retrocausality hypothesized so far, starting with the very first, the famous Wheeler and Feynman's 1949 one $\left [7\right ]$, never entered mainstream physics. However, we would think that the quantum computational speedup, a relatively recent discovery, offers an entirely new ground for addressing the retrocausality issue.

We go now to the derivation of the advanced knowledge rule by flanking the ordinary quantum description of Grover algorithm with its classical logic description.

By the way, for its special features, Grover's quantum search algorithm will play a special role in the present work:

\medskip

(i) In its simplest instance, and any instance in Long's version of it $\left [8\right ]$, it provides the solution with certainty, what simplifies things.

\medskip

(ii) The solution is an invertible function of the setting of the problem -- a condition for the reversibility of the quantum process in between: this will become important.

\medskip
\medskip

(iii) It is an optimal quantum algorithm, namely one that demonstrably solves its problem with the minimum possible number of quantum computation steps $\left [9\right ]$. This is obviously important when comparing the number of computation steps foreseen by the advanced knowledge rule in the optimal logical case with that taken by the quantum algorithm.

\medskip

(iv) It is a most universal quantum algorithm that can solve with quantum speedup any NP problem.

\medskip

Let us develop our classical logic description of what the quantum algorithm does first on the simplest instance of Grover algorithm. We will see that things immediately generalize to the full fledged Grover algorithm (Long's version of it), oracle quantum computing and, even beyond quantum computing, a most general form of quantum correlation (in quantum algorithms, that between the problem-setting and the corresponding solution of the problem).

The problem to be solved is as follows. Bob, the problem-setter, hides a ball in a chest of four drawers -- we call the number of the drawer with the ball the \textit{problem-setting}. Alice, the problem-solver, is to locate the ball by opening drawers. In the classical case, she has to open up to three drawers. If the ball is only in the third drawer opened, she has solved the problem; if it is not even in that drawer, it must be in the only drawer not yet opened and she has solved the problem of locating the ball as well. In the optimal quantum case -- by the optimal quantum search algorithm of Grover -- she always solves the problem by opening just one drawer (of course, for a quantum superposition of the four possible drawer numbers). Always solving the problem by opening just one drawer would obviously be impossible in the classical  case.

By the way, we will adopt a quantum description of Grover algorithm different from the usual one, which, as that of the other quantum algorithms, is limited to the process of solving the problem (where the quantum speedup resides) and the final measurement of the solution. It is therefore (trivially) incomplete. The canonically complete quantum description of a quantum process is: initial measurement, unitary evolution, and final measurement. We therefore complete the quantum description of Grover algorithm by adding to the quantum description of the process of computing and measuring the solution of the problem that of the upstream process of setting the problem.

Our ``canonically complete'' quantum description of the quantum algorithm is therefore as follows. An initial measurement of the problem-setting (e. g. the number of the drawer with the ball) in a quantum superposition (or, indifferently, mixture) of all the possible problem-settings, selects one of them at random\protect\footnote{
The problem-setter could unitarily change the problem-setting selected at random into any desired one. However, here, this would change nothing.
}. The unitary computation and measurement of the solution follow. 
We will call this quantum description of the quantum algorithm, extended to the process of setting the problem, its \textit{ordinary quantum description}, ``ordinary'' because it is still the causal quantum description of the evolutions of mainstream quantum mechanics.

By the way, further on we will have to further complete this ``canonically complete'' quantum description of the quantum algorithm, this time in a fundamental way.

We go now to the classical logic description of what the quantum algorithm does. The quantum fact of always solving the problem by opening a single drawer has a classical logic equivalent. It is as if the problem-solver always knew in advance, before beginning her problem-solving action, that the ball is in one of two given drawers (one of which, of course, the one with the ball) and took advantage of this knowledge to solve the problem by opening just one of them. Indeed, if the ball is in the drawer opened, she has solved the problem; if not, it must be in the other drawer and she has solved the problem of locating the ball as well, in either case having opened just one drawer. More precisely, what the problem-solver should know in advance is the mutually exclusive $or$ of all the possible ways of knowing in advance that the ball is in a pair of drawers. Note that this also means knowing in advance half of the information that specifies the solution of the problem she will produce and measure in the future.

By the way, note that we are developing a classical logic description of what the quantum algorithm does. We are not talking of the quantum algorithm as it were a classical algorithm. In other words, here, classical logic should not be confused with classical physics, what might be a temptation in the present case.

Let us go back to our line of thought. Always knowing in advance that the ball is in a pair of drawers is logically necessary to always solve the problem by opening just one drawer. Then the question becomes: from where does Alice, before beginning her problem-solving action, always get that information about the solution? Since, to her, the only possible source of it is the solution of the problem she will produce and measure in the future, the only possible answer is that it comes to her from her future measurement of the solution.

Then (``then'' in causal order), she can produce the solution benefiting of that advanced knowledge. We should keep in mind that we are under a classical logic equivalent of what the quantum algorithm does. In it, the needed number of computation steps is that logically needed. All this, of course, is in the mutually exclusive $or$ all the possible ways of knowing in advance half of the information about the solution. Eventually, note that this is exactly the advanced knowledge rule logically implied by the sheer existence of the present quantum speedup.

Moreover, providing the same quantum speedup given by Grover algorithm, besides being implied by its quantum speedup, the advanced knowledge rule implies it. In other words, the rule in question turns out to be just a classical logic interpretation of the present instance of Grover algorithm. We will soon see that both the rule and its being an interpretation extend to the full fledged Grover algorithm, oracle quantum computing, and a most general form of quantum correlation (in quantum algorithms, between the setting and the solution of the problem).

\subsection{
Completing the quantum description of Grover algorithm
}
Of course, there should be correspondence between the classical logic description of the quantum algorithm and its ordinary quantum description. However, its ordinary quantum description does not and, being causal in character, could not describe any causal loop. Coherently with the present line of thought, and despite the fact that it gives the correct number of computation steps needed to solve the problem (although in an unfereseeably different way for each quantum algorithm), we must conclude that it is incomplete. Naturally, it should be completed in a way that satisfies the correspondence in question.

In view of what will follow, note that the process between the outcome of the initial measurement and the state immediately before the final measurement is both unitary and time-reversible. Since the final measurement outcome coincides with the state immediately before the final measurement (which gives the solution with certainty), we will also speak, for short, of the unitary evolution between the initial and final measurement outcomes, even if it comprises a final measurement that does not change the quantum state. Also note that this unitary evolution (i.e. the computation of the solution) leaves the eigenvalue selected by the initial measurement, i.e. the problem-setting, unaltered.

We should now go and complete this ordinary quantum description in a way that brings it into accord with the classical logic description. It turns out that there is a most simple way of doing this. It is by time-symmetrizing it; we will see that it is also the only possible way.

Let us outline this operation. We will examine it in further detail in Section 2 and formalize it in Section 3.

The first time-symmetrization operation consists in evenly sharing between the initial measurement of the problem setting and final measurement of the corresponding solution the selection of the two. We should just postpone a suitable part of the projection of the quantum state associated with the initial measurement until the time of the final measurement. Note that this is possible since the unitary computation of the solution never changes the problem-setting. Since there is a plurality of possible ways of evenly sharing, at the end we should take their quantum superposition. In each time-symmetrization instance, time-symmetrically, the selection performed by the initial measurement should propagate forward in time by the unitary computation of the solution, that performed by the final measurement backward in time, by the inverse transformation. We should eventually take the quantum superposition of all the time-symmetrization instances.

As we will see, this time-symmetrization leaves the unitary part of the ordinary quantum description of the quantum algorithm (the computation of the solution) mathematically unaltered but changes the description of the behavior of causality along it. As seen by the problem-solver (we will explain this specification further below), the single causal process of the ordinary quantum description is replaced by the quantum superposition of the causal loops whose mutually exclusive $or$ is logically implied by the sheer existence of the quantum speedup. From now on, we will call the (physical) causal loops in question \textit{time-symmetric} because highlighted by a time-symmetrization procedure; we will call in the same way also the form of retrocausality entailed in them. 

By the way, note that the retrocausality we have been dealing with is, in the first place, logical in character: it is the one present in the classical logic description of what the quantum algorithm does. Of course, retrocausality can have a purely logical form, all the paradoxes originated by it have this form: think of the grandfather paradox. However, with the time-symmetrization of the ordinary quantum description of the quantum algorithm, this logical retrocausality becomes physical, quantum indeed (the quantum superposition of causal loops). 

Of course, the retrocausality made explicit by completing the ordinary quantum description of the quantum algorithm must also be implicit in its ordinary quantum description, namely the unitary interplay between quantum superpositions and interferences that characterizes the computation of the solution. A quantum algorithm is a rather complex process. Naturally, it would be desirable to know whether there is some more simple quantum phenomenon whose classical logic description entails the retrocausality we are dealing with.

Let us show that this is indeed the case, the phenomenon in question being \textit{quantum state superposition}. Besides being an essential ingredient of the quantum speedup (think of \textit{parallel quantum computation}), it is, of course, one of the pillars of quantum mechanics.

Let us recall the classical logic description of the prescientific notion of state superposition: the monk being on the two banks of the river at the same time. Since the two possibilities are logically mutually exclusive with one another, this is apparently a logical impossibility. However, it is an impossibility that vanishes by resorting to retrocausality. After reaching one bank, the monk should sufficiently go backward in time by the inverse of the path followed to reach that bank and then forward in time again but on a different path that leads him to the other bank.

Going now to the quantum world, under the assumption that quantum algorithms can be described in a classical logic way, also the quantum superpositions that are an essential part of them should be describable in this way. Let us see how the above classical logic description of the monk being on the two banks of the river at the same time can be applied, for example, to a photon passing in quantum superposition along the two branches of a Mach-Zehnder interferometer. 

The two branches should be tuned in such a way that the photon always ends up in the same detector. In this way, the whole process until ``reaching'' the detector (we mean: immediately before detection) is unitary and can therefore be run backward in time by the inverse transformation. Of course, there is a classical logic description of the photon (with its phase) ``reaching'' the detector having run either branch but these two possibilities are mutually exclusive with one another. However, there can be a classical logic description also in this case provided that one opens to retrocausality. After ``reaching'' the detector (immediately before detection) having gone along one of the two branches, the photon should go backward in time, by the inverse of the time-forward unitary transformation, until before entering the Mack-Zehnder interferometer; then, going forward in time again, it should reach the same detector through the other branch. This time it can also be detected.

This is a classical logic description of the photon going in quantum superposition through the upper and lower branch. One can see that any such description necessarily involves retrocausality.

Going now to the unitary quantum computation of the solution, this is a sequence of generations of quantum state superpositions and interferences. The former can thus be described in a classical logic way by resorting to the causality zigzags we are dealing with, the latter, of course, without any zigzag. After the simplifications, we should turn out with the advanced knowledge rule derived in a much more compact way by the classical logic description of the resulting quantum speedup. Now, about the advantage of relating the quantum speedup to retrocausality rather than quantum superposition, it is, of course, that of moving from a qualitative to a quantitative explanation of it.

Summing up, at the origin of the present classical logic description of the quantum speedup there is the simple fact that the classical logic description of a quantum state superposition necessarily entails retrocausality. It is, first, a logical retrocausality and then, after completing the quantum description of the quantum algorithm by time-symmetrizing its ordinary quantum description, a physical one, of course in the quantum world.

\subsection{
Generalization
}

We will see that the reason for time-symmetrizing the ordinary quantum description of Grover algorithm generalizes to that of any unitary evolution between two one-to-one correlated measurement outcomes that does not change the eigenvalue of the observable selected by the initial measurement (in quantum algorithms, the problem-setting selected by the initial measurement is not changed by the unitary computation of the solution).

The cases are two.

One is that the unitary evolution builds up one-to-one correlation between two initially unrelated (commuting) observables without changing the eigenvalue of the observable selected by the initial measurement. This building up can always occur with quantum speedup. This is, of course, the case of quantum algorithms but the unitary evolution between the two one-to-one correlated measurement outcomes may not be a quantum algorithm at all. In any case, that building up can always occur with (quadratic) quantum speedup. When we are not dealing with a quantum algorithm, obviously, the notion of problem solver and her advanced knowledge of half of the information about the solution vanishes. What instead survives is the reduction of the dimension of the Hilbert space in which the search for one-to-one correlation takes place. It is this search what can occur with quantum speedup.

The other case is that there is no such building up because the two observables are already maximally entangled before the beginning of the quantum process. This is, for example, the case of quantum nonlocality.

Consider the measurement (in the same basis) of two maximally entangled observables in a state of spatial separation between the two respective subsystems. Of course, there is one-to-one correlation between the two measurement outcomes and the unitary process of spatial separation does not change the eigenvalue selected by the initial measurement. As shown in $\left [5\right ]$, time-symmetrizing the ordinary quantum description of quantum nonlocality (exactly as we did with that of quantum algorithms) yields a quantum superposition of causal loops that play the role of the hidden variables envisaged by Einstein and the others in their celebrated EPR paper $\left [10\right ]$. Indeed, from the outcome of the first performed measurement (on one of the two subsystems) they locally go backward in time to the time the two subsystems were not yet space separate, where they locally change the state of the other subsystem in a way that ensures the appropriate correlation between the two future measurement outcomes. In other words, the causal loops we are dealing with exactly play the role of, or even be, the hidden variables envisaged by Einstein and the others.

Note the many commonalities between the quantum speedup and quantum nonlocality that there are under the present theory, given that both do something that is impossible to do in the classical case, respectively solving a problem with fewer computation steps than classically possible and instantaneously changing the state of an object at a distance:

\medskip

 (i) both do so by availing themselves of the present time-symmetric form of retrocausality,

\medskip

(ii) in both cases, the classical logic description of the quantum process highlights the incompleteness of its ordinary quantum description,

\medskip

(iii) in both cases, the ordinary quantum description is completed by time-symmetrizing it, and

\medskip

(iv) in both cases we are dealing with a retrocausality internal to the quantum process that cannot be used by observers external to it, for example to communicate backward in time or instantaneously at a distance.

\medskip

By the way, it would be another time that Einstein saw further than others, even further than has been thought so far.

As we will see further on, Costa de Beauregard $\left [11\right ]$, in 1953, had already provided a local, retrocausal, explanation of quantum nonlocality very similar to the present one but, of course, without the possibility of exploiting the time-symmetric form of retrocausality presently shown to be implicit in both quantum algorithms and quantum nonlocality.

We would like to conclude this introduction by recalling the priorities along the path that led to the discovery of the quantum computational speedup, a discovery that, under the present theory, would even be more fundamental than has been thought so far (only the priorities, of course followed by the well-known subsequent developments). In 1969, Finkelstein showed that computation is possible in the quantum world and introduced the fundamental notion of quantum bit $\left [12\right ]$. In 1982, Feynman highlighted the fundamentally higher computational efficiency of the quantum world $\left [13\right ]$. In 1985, Deutsch devised the seminal quantum algorithm that delivers a quantum computational speedup $\left [14\right ]$.

The reader not interested in further technical detail or the formalizations of what exposed until now can go to Subsection 4.2. ``Looking forward''.

\section{
Addenda
}
Before moving on to the formalizations of Section 3., let us open some of the ``we will see'' put for compactness in the previous section.

First, we extend to any number, $N$, of drawers the demonstration that the quantum speedup of Grover quantum search algorithm implies the advanced knowledge rule.

Grover algorithm yields a \textit{quadratic} quantum speedup, that is, it always solves the problem by opening $\ensuremath{\operatorname*{O}}\left (\sqrt{N}\right )$ drawers against the $\ensuremath{\operatorname*{O}}\left (N\right )$ of the best classical case\protect\footnote{
The four drawer instance of Grover algorithm gives the solution with certainty; with more than four drawers, by ``Grover algorithm'' we will always mean Long's version of it $\left [8\right ]$, which always yields the solution with certainty.
}. Also now, always solving the problem opening $\ensuremath{\operatorname*{O}}\left (\sqrt{N}\right )$ drawers is logically equivalent to the fact that Alice always knows in advance (in a mutually exclusive way) one of the possible halves of the information about the number of the drawer with the ball, namely that the ball is (in a mutually exclusive way) in one of the possible tuples  of $\sqrt{N}$ drawers. She can therefore use this knowledge to logically produce the solution by opening just $\ensuremath{\operatorname*{O}}\left (\sqrt{N}\right )$ drawers. This is of course the advanced knowledge rule for the full fledged Grover algorithm.

We go now to the completion of the ordinary quantum description of the quantum algorithm by its time-symmetrization. Particular attention will be paid to showing that all the operations performed in it are both (i) physically legitimate and (ii) necessary to bring the quantum description of the quantum algorithm into accord with its classical logic description. We are in uncharted waters and checking consistency is obviously important.

Let us recall, for convenience, our ``ordinary quantum description of quantum algorithms''.

The initial measurement of the problem-setting in a quantum superposition (or mixture) of all the possible problem settings selects one of them at random; this is followed by the unitary computation of the corresponding solution (which never changes the problem-setting selected by the initial measurement) and the final measurement of the latter. The solution is an invertible function of the problem-setting and is determined with cetainty.

 Note that the process that connects the initial with the final measurement outcome is unitary and thus time-reversible. As already mentioned, for shortness, we ``confuse'' the final measurement outcome with the state immediately before it (which coincides with it).

We go now to describe the time-symmetrization of this ordinary quantum description; we will discuss it next.

The basic assumption is that the selection of the problem setting and the corresponding solution (one an invertible function of the other) evenly shares between the initial measurement of the problem setting and the final measurement of the solution, in a quantum superposition of all the possible ways of evenly sharing. In each time-symmetrization instance, the selection performed by the initial measurement propagates forward in time, that performed by the final measurement backward in time.

 Going into further detail, in each time-symmetrization instance, the initial measurement of the problem-setting in a quantum superposition of all the possible problem-settings randomly selects one of the possible halves of the information abut it, therefore selecting the quantum superposition of a number of problem-settings that is the square root of the total number. The successive unitary evolution generates a quantum superposition of tensor products, each one of the problem-settings selected by the initial measurement tensor product the corresponding solution. The final measurement of the solution randomly selects a single tensor product. Its measurement outcome propagates backward in time by the inverse of the time-forward unitary transformation. We should eventually take the quantum superposition of all the possible time-symmetrization instances.

Let us discuss now the legitimacy of the above operations.

First, since the unitary computation of the solution never changes the problem-setting selected by the initial measurement, postponing the associated projection of the quantum state, or any part of it, along the subsequent unitary evolution is a legitimate operation.

We go now to the backward in time propagation of the final measurement outcome. Of course, it is an unusual operation. However, it is mathematically legitimate here. Indeed, the final measurement outcome -- a single problem-setting tensor product the corresponding solution -- coincides with the final measurement outcome of the ordinary quantum description. This given, the backward in time propagation of the final measurement outcome by the inverse of the unitary transformation of the ordinary quantum description just undoes the postponing of the projection associated with the initial measurement. Indeed, it rebuilds the mathematics of the ordinary quantum description of the quantum algorithm.

Of course, the legitimacy of the above operations does not justify the fact of assuming them. We will soon see that this is necessary to complete the quantum description of the quantum algorithm in a way that corresponds to its classical logic description.

Now, let us look at things as a whole:

\medskip

(i) each time-symmetrization instance consists of a time-forward propagation (by a unitary transformation) going from the outcome of the initial measurement to the state immediately before the final measurement and a time-backward propagation (by the inverse of the previous unitary transformation) going from the outcome of the final measurement to exactly the beginning of the problem-solver's action, where it changes her complete ignorance of the solution into the knowledge of one of the possible halves of the information about it. One can see that this is a causal loop, an effective one that changes a past quantum state (we will see this formally in Subsection 3.1.).

\medskip

(ii) the  backward in time propagation of each time-symmetrization instance, which inherits both selections, is an instance of the time-symmetrized  quantum algorithm. Since all backward in time propagations are the same (mathematically, the unitary part of the ordinary quantum description), any one of them is their quantum superposition, thus the time-symmetrized quantum algorithm.

\medskip

Summing up, the time-symmetrization we are dealing with completes the ordinary (causal) quantum description of the quantum algorithm by:

\medskip

(a) Leaving the unitary part of the ordinary quantum description mathematically unaltered.

\medskip

(b) Changing the description of the behavior of causality along that part. As viewed by the problem-solver, the single causal process of the ordinary quantum description is replaced by the quantum superposition of the causal loops whose mutually exclusive $or$ is logically implied by the sheer existence of the quantum speedup.

\medskip

Let us explain the reason for the specification ``as viewed by the problem-solver''.

In the first place, the causal loops logically implied by the sheer existence of the quantum speedup, which tell the problem-solver half of the information about the solution, are, of course, with respect to her.

Now, the ordinary quantum description of the quantum algorithm is instead with respect to the customary observer external to the quantum process. It could not be with respect to the problem-solver: it would tell her, before she begins her problem-solving action, the problem-setting selected by the initial measurement and thus the solution of the problem (here, they are both the number of the drawer with the ball). To have the quantum description with respect to the problem-solver (\textit{relativized} to her), we should conceal from her the problem-setting selected by the initial measurement (see the formalization of Subsection 3.1.2. for further detail).

Similarly, we should relativize with respect to the problem-solver the causal loops generated by time-symmetrizing the ordinary quantum description of the quantum algorithm, by always concealing from her the selection performed by the initial measurement (see the formalization of Subsection 3.1.4).

This is the reason for the above specification: ``as viewed by the problem-solver''.

Also note that the time-symmetrization we are dealing with, besides consisting of legitimate operations, is the only possible way of completing the ordinary quantum description of the quantum algorithm in a way that describes the causal loops logically implied by the sheer existence of its quantum speedup. 

In fact, with the kind of process we are dealing with, to describe a causal loop (any one), we must properly share the selection of the problem-setting between initial and final measurements, with the part of the selection performed by the initial measurement propagating forward in time by the unitary computation of the solution and that performed by the final measurement backward in time by the inverse of that unitary computation. To obtain exactly the causal loops whose mutually exclusive $or$ is logically implied by the sheer existence of the quantum speedup, this sharing must be even. 

The above exactly identifies the time-symmetrization we have performed. It also shows that this time-symmetrization is mandatory: it is the only way of generating the quantum superposition of causal loops that must correspond physically to the classical logic description of the quantum algorithm.

\section{
Formalization
}
This section is divided into two parts. In the first, we formalize the notion of completing the ordinary quantum description of the quantum algorithm by time-symmetrizing it. The second is the generalization from the quantum algorithm to quantum correlations.

\subsection{
Completing the ordinary quantum description of quantum algorithms by time-symmetrizing it
}
The completion of the quantum description of the quantum algorithm consists of the following steps:

\medskip

(1) Extend the usual quantum description of the quantum algorithm, limited to the process of solving the problem, to the upstream process of setting the problem. We will work on the simplest, four-drawer instance of Grover's quantum search algorithm and then generalize.

\medskip

(2) The ordinary quantum description of the quantum algorithm is with respect to the customary observer external to the quantum process. We show how to relativize it with respect to the problem solver, to whom the problem setting selected by the initial measure is to be kept hidden.

\medskip

\medskip (3) Complete the ordinary quantum description of the quantum algorithm by time-symmetrizing it.

\medskip

(4) Relativize with respect to the problem-solver the causal loops generated by time-symmetrizing the ordinary quantum description of the quantum algorithm, which are of course with respect to the customary external observer. Note that the causal loops whose mutually exclusive $or$ is logically implied by the sheer existence of the quantum speedup, each telling the problem-solver in advance one of the possible halves of the information about the solution, are of course with respect to her. 

\medskip

The above steps (1) to (4) are developed in the following subsections.

\medskip

\subsubsection{Extending the usual quantum description of quantum algorithms to the process of setting the problem}
The usual quantum description of quantum algorithms is limited to the unitary process of solving the problem, where the quantum speedup resides, and the final measurement of the solution. It is trivially incomplete since it lacks the initial measurement. Of course, the canonical quantum description of a quantum process must consist of initial measurement, unitary evolution, and final measurement. We extend the quantum description of the process of solving the problem to the upstream process of setting the problem. An initial measurement of the problem-setting in a quantum superposition (or, indifferently, mixture) of all the possible problem-settings selects one of them at random. This is followed by the unitary evolution that computes the corresponding solution and the final measurement of the latter.

Let us introduce the notation. We make reference to the simplest, four drawer, instance of Grover algorithm:

\medskip

We number the four draws in binary notation: $00 ,01 ,10 ,11$.  

\medskip

We will need two quantum registers of two \textit{quantum bit}s each:

\medskip

A register $B$ under the control of the problem-setter Bob contains the \textit{problem-setting} (the number of the drawer in which the ball is hidden). The content of register $B$ is \textit{the problem-setting observable
$\hat{B}$   }of eigenstates and corresponding eigenvalues:

\medskip \medskip

$\vert 00 \rangle _{B}$ and $00$; $\vert 01 \rangle _{B}$ and $01$; $\vert 10 \rangle _{B}$ and $10$; $\vert 11 \rangle _{B}$ and $11$. 

\medskip 

              A register $A$ under the control of the problem-solver Alice contains the number of the drawer opened by her. The content of register $A$ is the observable
$\hat{A}$  of eigenstates and corresponding eigenvalues:

\medskip

$\vert 00 \rangle _{A}$ and $00$; $\vert 01 \rangle _{A}$ and $01$; $\vert 10 \rangle _{A}$ and $10$; $\vert 11 \rangle _{A}$ and $11$.

\medskip

Measuring
$\hat{A}$  at the end of the quantum algorithm yields the number of the drawer with the ball selected by the initial measurement, namely the solution of the problem. Note that
$\hat{B}$  and
$\hat{A}$  commute.

\medskip

The quantum description of the extended quantum algorithm, namely its \textit{ordinary quantum description} is: 

\medskip

\begin{equation}\begin{array}{ccc}\begin{array}{c}\;\text{time }t_{1}\text{, meas. of}\;\hat{B}\end{array} & t_{1} \rightarrow t_{2} & \text{time }t_{2}\text{, meas. of}\;\hat{A} \\
\, & \, & \, \\
\begin{array}{c}\left (\vert 00 \rangle _{B} +\vert  01 \rangle _{B} +\vert 10 \rangle _{B} +\vert  11 \rangle _{B}\right ) \\
\left (\vert 00 \rangle _{A} +\vert 01 \rangle _{A} +\vert 10 \rangle _{A} +\vert 11 \rangle _{A}\right )\end{array} & \, & \, \\
\Downarrow  & \, & \, \\
\vert 01 \rangle _{B}\left (\vert 00 \rangle _{A} +\vert 01 \rangle _{A} +\vert 10 \rangle _{A} +\vert 11 \rangle _{A}\right ) &  \Rightarrow \hat{\mathbf{U}}_{1 ,2} \Rightarrow  & \vert 01 \rangle _{B}\vert  01 \rangle _{A}\end{array}\tag{Table I}
\end{equation}

\medskip

Here and in the following we disregard normalization.

We should start from the quantum state in the upper-left corner of Table I diagram and then follow the vertical and horizontal arrows. Initially, at time $t_{1}$, both registers, each by itself, are in a quantum superposition of all their basis vectors. By measuring
$\hat{B}$  (the content of $B$) in this initial state, Bob selects the number of the drawer with the ball at random, say it comes out $01$ -- see the state of register $B$ under the vertical arrow\protect\footnote{
By the way, Bob would be free to unitarily change at will the sorted out number of the drawer with the ball. We omit this operation that would change nothing here.
}.

It is now Alice's turn. The unitary part of her problem-solving action -- by the unitary transformation $\hat{\mathbf{U}}_{1 ,2}$ -- produces the solution of the problem, of course, never changing the problem-setting. 

Note that we will never need to know $\hat{\mathbf{U}}_{1 ,2}$ (i.e. Grover algorithm). We only need to know that there can be a unitary transformation between the outcome of the initial measurement and that of the final measurement, what is always the case since the final measurement outcome is an invertible function of the initial one.

At the end of Alice's problem-solving action, at time $t_{2}$, register $A$ contains the solution of the problem, i.e. the number of the drawer with the ball -- bottom-right corner of the diagram. 

Eventually, by measuring
$\hat{A}$  (the content of $A$), which is already in one of its eigenstates, Alice acquires the solution with probability $1$ -- the quantum state in the bottom-right corner of the diagram naturally remains unaltered throughout this measurement. Note that, as a consequence, the process between the initial and the final measurement outcomes is unitary, and thus time-reversible, despite comprising the final measurement.

\subsubsection{
Relativizing the ordinary quantum description of the quantum algorithm with respect to the problem-solver
}

The ordinary quantum description of the quantum algorithm of Table I is with respect to the customary external observer, that is, an observer external to the quantum process. It is not a description for the problem-solver Alice. In fact, it would tell her, immediately after the initial measurement of
$\hat{B}$, the problem-setting selected by the initial measurement and thus the solution of the problem. By the way, here they are both the number of the drawer with the ball. Of course, in the quantum description of the quantum algorithm with respect to Alice, she must be concealed from the outcome of the initial measurement.

Since Alice's problem-solving action never changes the number of the drawer with the ball selected by the initial measurement and the observables
$\hat{B}$ and
$\hat{A}$  commute, describing this concealment is simple. It suffices to postpone   after the end of Alice's problem-solving action the projection of the quantum state associated with the initial measurement of
$\hat{B}$  (or, indifferently, to postpone the very measurement of
$\hat{B}$). The result is the following quantum description of the quantum algorithm with respect to Alice:

\medskip

\begin{equation}\begin{array}{ccc}\begin{array}{c}\text{time}\text{ }t_{1}\text{, meas. of}\;\hat{B}\end{array} & t_{1} \rightarrow t_{2} & \text{\text{time }\text{}}t_{2}\text{, meas. of}\;\hat{A} \\
\, & \, & \, \\
\begin{array}{c}\left (\vert 00 \rangle _{B} +\vert  01 \rangle _{B} +\vert 10 \rangle _{B} +\vert  11 \rangle _{B}\right ) \\
\left (\vert 00 \rangle _{A} +\vert 01 \rangle _{A} +\vert 10 \rangle _{A} +\vert 11 \rangle _{A}\right )\end{array} &  \Rightarrow \hat{\mathbf{U}}_{1 ,2} \Rightarrow  & \begin{array}{c}\vert 00 \rangle _{B}\vert  00 \rangle _{A} +\vert 01 \rangle _{B}\vert  01 \rangle _{A} + \\
\vert 10 \rangle _{B}\vert  10 \rangle _{A} +\vert 11 \rangle _{B}\vert  11 \rangle _{A}\end{array} \\
\, & \, & \Downarrow  \\
\, & \, & \vert 01 \rangle _{B}\vert  01 \rangle _{A}\end{array}\tag{Table II}
\end{equation}

\medskip 

 To Alice, the initial measurement of \textbf{$\hat{B}$} leaves the initial state of register $B$ unaltered. At the beginning of her problem-solving action she is completely ignorant of the number of the drawer with the ball selected by Bob -- see the state of register $B$ on the left of $ \Rightarrow \hat{\mathbf{U}}_{1 ,2} \Rightarrow $.

 The unitary transformation   $\hat{\mathbf{U}}_{1 ,2}$ is then performed for a quantum superposition of the four possible numbers of the drawer with the ball. The state at the end of it -- on the right of the horizontal arrows -- is a quantum superposition of tensor products, each a possible number of the drawer with the ball in register $B$ tensor product the corresponding solution (that same number) in register $A$. Since
$\hat{B}$ and
$\hat{A}$  commute, the final measurement of
$\hat{A}$  must project this superposition on the tensor product of the number of the drawer with the ball initially selected by Bob and the corresponding solution -- bottom-right corner of Table II diagram.

Note that this relativization relies on relational quantum mechanics $\left [15 ,16\right ]$ where the quantum state is relative to the observer (also note that, here, one observer is external and the other internal to the quantum process). In the following, we will continue to call the quantum description with respect to the external observer \textit{the ordinary quantum description of the quantum algorithm}. Instead we will always specify that the other quantum description is that with respect to the problem-solver.

\subsubsection{
Time-symmetrizing the ordinary quantum description of the quantum algorithm
}
As anticipated in the Rationale, in order to obtain the description of the quantum superposition of the causal loops whose mutually exclusive $or$ is implied by the sheer existence of an optimal quantum speedup, we should first time-symmetrize the ordinary quantum description of the quantum algorithm, which is with respect to the customary external observer, then relativize the causal loops obtained with respect to the problem-solver.

To start with, we should require that the selection of the information that specifies the sorted out pair of correlated measurement outcomes among all the possible pairs evenly shares between the initial and final measurements. The half information selected by the initial measurement should propagate forward in time, by  $\hat{\mathbf{U}}_{1 ,2}$ until becoming the state immediately before the final measurement. Time-symmetrically, the complementary selection performed by the final measurement should propagate backward in time, by
$\hat{\mathbf{U}}_{1 ,2}^{\dag }$, until replacing the previous outcome of the initial measurement.

 Note that the selection, by the initial measurement, of half of the information that specifies the problem-setting is made possible by two facts: (i) the measurement occurs in a quantum superposition of all the possible problem-setting, therefore it is sufficient to reduce it to the measurement of one of the possible halves of the bits of the problem-setting in question (in register $B$) and (ii) the problem-setting never changes along the unitary evolution between the two measurement outcomes, therefore the measurement of the other half of the bits can be postponed along that evolution until making it simultaneous with the final measurement.  Of course, as the even sharing of the selection can be performed in many possible mutually exclusive ways, eventually we should take their quantum superposition.

We consider the time-symmetrization instance where the initial measurement of
$\hat{B}$  reduces to the measurement of
$\hat{B}_{l}$, the left bit of the two-bit number contained in register $B$; say that it randomly selects left bit $0$. The final measurement of
$\hat{A}$  should correspondingly reduce to that of
$\hat{A_{r}}$  the right bit of the two-bit number contained in register $A$; say that it randomly selects right bit $1$. The randomly selected number of the drawer with the ball is thus $01$.

In this particular time-symmetrization instance, the ordinary quantum description of the quantum algorithm with respect to the external observer
of Table I becomes:

\medskip

\begin{equation}\begin{array}{ccc}\begin{array}{c}\;\text{time }t_{1}\text{, }\text{}\text{meas. of}\;\hat{B}_{l}\end{array} & t_{1 \rightleftarrows }t_{2} & \text{time }\text{}t_{2}\text{, meas. of}\;\hat{A_{r}} \\
\, & \, & \, \\
\begin{array}{c}\left (\vert 00 \rangle _{B} +\vert  01 \rangle _{B} +\vert 10 \rangle _{B} +\vert  11 \rangle _{B}\right ) \\
\left (\vert 00 \rangle _{A} +\vert 01 \rangle _{A} +\vert 10 \rangle _{A} +\vert 11 \rangle _{A}\right )\end{array} & \, & \, \\
\Downarrow  & \, & \, \\
\begin{array}{c}\left (\vert 00 \rangle _{B} +\vert  01 \rangle _{B}\right ) \\
\left (\vert 00 \rangle _{A} +\vert 01 \rangle _{A} +\vert 10 \rangle _{A} +\vert 11 \rangle _{A}\right )\end{array} &  \Rightarrow \hat{\mathbf{U}}_{1 ,2} \Rightarrow  & \vert 00 \rangle _{B}\vert  00 \rangle _{A} +\vert 01 \rangle _{B}\vert  01 \rangle _{A} \\
\, & \, & \Downarrow  \\
\vert 01 \rangle _{B}\left (\vert 00 \rangle _{A} +\vert 01 \rangle _{A} +\vert 10 \rangle _{A} +\vert 11 \rangle _{A}\right ) &  \Leftarrow \hat{\mathbf{U}}_{1 ,2}^{\dag } \Leftarrow  & \vert 01 \rangle _{B}\vert  01 \rangle _{A}\end{array}\tag{Table III}
\end{equation}

\medskip

 The initial measurement of $\hat{B}_{l}$, selecting the
$0$
of
$01$, projects the initial quantum superposition of all the basis vectors of register $B$ on the superposition of those
beginning with
$0$
-- vertical arrow on the left of the diagram.  The latter superposition causally propagates forward in time, by
$\hat{\mathbf{U}}_{1 ,2}$, into the superposition of the two tensor products on the right of the right looking horizontal arrows. Then the
final measurement of
$\hat{A_{r}}$, selecting the
$1$
of
$01$, projects the latter superposition on the term ending with $1$ under the right vertical arrow. Note that this outcome of the final measurement, which inherits both selections, coincides with that of the ordinary quantum description of the quantum algorithm -- see the bottom-right corner of Table I diagram. Time-symmetrically, this final measurement outcome causally propagates backward in time, by
$\hat{\mathbf{U}}_{1 ,2}^{\dag }$, until becoming the definitive outcome of the initial measurement. This backward in time propagation (which inherits both selections) is an instance of the time-symmetrization of the ordinary quantum description
of the quantum algorithm -- bottom line of the diagram. Note that, mathematically (disregarding the time-direction of causality) it is the ordinary quantum description of the unitary part of the quantum algorithm back again -- see the bottom line of the diagram in Table I. Indeed, the symbol ``$ \Leftarrow \hat{\mathbf{U}}_{1 ,2}^{\dag } \Leftarrow $'' is mathematically equivalent to ``$ \Rightarrow \hat{\mathbf{U}}_{1 ,2} \Rightarrow $''.

Also note that, together, the forward and backward in time propagations form a causal loop that changes its own past;  the outcome of its initial measurement of
$\hat{B}_{l}$  is changed into the outcome of the initial measurement of
$\hat{B}$  of the ordinary quantum description; compare the state of register $B$ under the left vertical arrow with the state of register $B$ in the bottom-left corner of the diagram.

We should eventually take the quantum superposition of all the time-symmetrization instances. We can see that, no matter how we evenly share the selection between initial and final measurements, the final measurement outcome, which inherits both selections, is always that of the final measurement of
$\hat{A}$  of the ordinary quantum description of the quantum algorithm. Correspondingly, its subsequent backward in time propagation by
$\hat{\mathbf{U}}_{1 ,2}^{\dag }$ (bottom line of the diagram), mathematically (disregarding the time-direction of causality), is always the ordinary quantum description of the unitary process between the two one-to-one correlated measurement outcomes, namely the bottom line of Table I diagram.

Summing up, mathematically, time-symmetrization leaves the ordinary quantum description of the unitary evolution between the two measurement outcomes unaltered but changes the structure of causality along this evolution. The single causal process of the ordinary quantum description is replaced by a quantum superposition of causal loops.

\subsubsection{
Relativizing the causal loops with respect to the problem-solver
}
The causal loops generated by time-symmetrizing the ordinary quantum description of the quantum algorithm, which is with respect to the external observer, do not yet  map on those logically implied by the sheer existence of the quantum speedup, which are with respect to the problem-solver. Indeed, we should relativize the causal loops of Table III with respect to the problem-solver. We should just conceal from her the part of the problem-setting selected by the initial measurement. This is done by postponing after her problem-solving action  the projection of the quantum state due to the initial measurement   of
$\hat{B}_{l}$. This yields the causal loop in the following table:

\medskip

\begin{equation}\begin{array}{ccc}\begin{array}{c}\;\text{time }t_{1}\text{, meas. of}\;\hat{B}_{l}\end{array} & t_{1 \rightleftarrows }t_{2} & \text{time }t_{2}\text{,meas. of}\;\hat{A_{r}} \\
\, & \, & \, \\
\begin{array}{c}\left (\vert 00 \rangle _{B} +\vert  01 \rangle _{B} +\vert 10 \rangle _{B} +\vert  11 \rangle _{B}\right ) \\
\left (\vert 00 \rangle _{A} +\vert 01 \rangle _{A} +\vert 10 \rangle _{A} +\vert 11 \rangle _{A}\right )  \end{array} &  \Rightarrow \hat{\mathbf{U}}_{1 ,2} \Rightarrow  & \begin{array}{c}\vert 00 \rangle _{B}\vert  00 \rangle _{A} +\vert 01 \rangle _{B}\vert  01 \rangle _{A} + \\
\vert 10 \rangle _{B}\vert  10 \rangle _{A} +\vert 11 \rangle _{B}\vert  11 \rangle _{A}\end{array} \\
\, & \, & \Downarrow  \\
\begin{array}{c}\left (\vert 01 \rangle _{B} +\vert  11 \rangle _{B}\right ) \\
\left (\vert 00 \rangle _{A} +\vert 01 \rangle _{A} +\vert 10 \rangle _{A} +\vert 11 \rangle _{A}\right )\end{array} &  \Leftarrow \hat{\mathbf{U}}_{1 ,2}^{\dag } \Leftarrow  & \vert 01 \rangle _{B}\vert  01 \rangle _{A} +\vert 11 \rangle _{B}\vert  11 \rangle _{A}\end{array}\tag{Table IV}
\end{equation}

\medskip

 The projection of the initial state of register $B$ due to the measurement of
$\hat{B}_{l}$  -- top-left corner of the diagram -- is postponed after Alice's problem-solving action, namely outside the present table which is limited to that action. With this postponement, the initial state of register $B$ goes unaltered through the initial measurement of
$\hat{B}_{l}$  becoming, by $\hat{\mathbf{U}}_{1 ,2}$, the quantum superposition of tensor products above the vertical arrow. Alice's final measurement of
$\hat{A_{r}}$, which in this time-symmetrization instance selects the value $1$ of the right bit of register $A$, projects the state above the vertical arrow on the state below it. Then (in causal order), this final measurement  outcome propagates backward in time by $\hat{\mathbf{U}}_{1 ,2}^{\dag }$ until becoming the definitive outcome of the initial measurement. This backward in time propagation is an instance of the time-symmetrized description of the quantum algorithm with respect to the problem-solver. It tells her, before she begins her problem-solving action, one of the possible halves of the information about the solution.

Note that we would have obtained the identical diagram by time-symmetrizing the description of the quantum algorithm relativized to the problem-solver of Table II.

Also note that, for the mathematical equivalence between $ \Leftarrow \hat{\mathbf{U}}_{1 ,2}^{\dag } \Leftarrow $ and $ \Rightarrow \hat{\mathbf{U}}_{1 ,2} \Rightarrow $, the  bottom line of Table IV diagram (mathematically) can also be read in the usual left to right way:

\medskip

\begin{equation}\begin{array}{ccc}\text{time }t_{1} & t_{1} \rightarrow t_{2} & \text{time }t_{2} \\
\, & \, & \, \\
\begin{array}{c}(\vert 01 \rangle _{B} +\vert  11 \rangle _{B}) \\
\left (\vert 00 \rangle _{A} +\vert 01 \rangle _{A} +\vert 10 \rangle _{A} +\vert 11 \rangle _{A}\right )  \end{array} &  \Rightarrow \hat{\mathbf{U}}_{1 ,2} \Rightarrow  & \vert 01 \rangle _{B} \vert 01 \rangle _{A} +\vert  11 \rangle _{B}\vert 11 \rangle _{A}\end{array}\tag{Table V}
\end{equation}

\medskip

Summing up, the entire causal loop is the zigzag diagram of table IV. We can see that it is one of the causal loops whose mutually exclusive $or$ is logically implied by the sheer existence of the quantum speedup. The state of register $B$  in the top-left part of the diagram tells us that Alice, before beginning her problem-solving action, is completely ignorant of the number of the drawer with the ball. Through the zig-zag, this state changes into the state of register $B$ in the bottom-left part of the diagram, which tells us that Alice, before beginning her problem-solving action, knows in advance that the ball is in drawer either $01$ or $11$. She can then solve the problem in an optimal logical way by opening just one of the two drawers, indeed as in the classical logic description of what the quantum algorithms does. In equivalent terms, the computational complexity of the problem to be solved by Alice quadratically reduces from the problem of locating the ball in one of $4$ drawers to that of locating it in $2 =\sqrt{4}$ drawers. In still equivalent terms, the dimension of the Hilbert space in which the search for the solution takes place quadratically reduces from $4$ basis vectors to $2 =\sqrt{4}$ basis vectors.

\subsection{
Generalization
}
Until now we have focused on the simplest instance of Grover algorithm, but the results obtained go well beyond it.

First, everything immediately generalizes to any number, $N$, of drawers: it suffices to increase from $2 =\sqrt{4}$ to $\sqrt{N}$ the number of quantum bits in each quantum register. We can see that all considerations remain unaltered: the problem-solver always knows in advance one of the possible halves of the information about the solution and can use this knowledge to produce the solution with correspondingly fewer computation steps, $\ensuremath{\operatorname*{O}}\left (\sqrt{N}\right )$ in the optimal logical and quantum case. In equivalent terms, the dimension of the  Hilbert space in which the search for the solution -- or one-to-one correlation between the two observables -- takes place reduces from $N$ basis vectors to $\sqrt{N}$ basis vectors.

Second, we never needed to look inside
$\hat{\mathbf{U}}_{1 ,2}$. This unitary transformation is just defined by the fact that it must build up one-to-one correlation between two initially unrelated, commuting, observables ($\hat{B}$ and
$\hat{A}$); this, without changing the eigenvalue of the observable selected by the initial measurement. Besides Grover algorithm, we could be dealing with any quantum algorithm where the solution is an invertible function of the problem-setting\protect\footnote{
Note that this condition is easily removed; it suffices that, at the end, the problem-setter first measures the problem-setting, then the solution.
} and is determined with certainty. In any such quantum algorithm, the problem-solver always knows in advance half of the information about the solution she will produce and measure in the future and can use this knowledge to produce the solution with fewer computation steps. This naturally includes all oracle problems.

Things generalize also beyond quantum algorithms. Let us consider a unitary evolution between an initial and a final one-to-one correlated measurement outcomes that does not change the eigenvalue of the observable selected by the initial measurement. In this kind of process, one is free to postpone the initial measurement at will. When made simultaneous with the final measurement, the time-symmetrization we are dealing with must already be in place to become the perfect atemporal symmetry between the simultaneous measurements of two maximally entangled observables -- see the states above and below the vertical arrow in Table II (thinking that the measurement of $\hat{B}$ is postponed and made simultaneous with that of
$\hat{A}$). This is for the principle of sufficient reason: there would be no reason for the prevalence of either measurement in the simultaneous selection of the pair of one-to-one correlated measurement outcomes.

Summing up, we can state that:

\medskip

\textit{The ordinary quantum description of any unitary evolution between an initial and a final one-to-one correlated measurement outcomes that does not change the eigenvalue of the observable selected by the initial measurement is incomplete and is completed by time-symmetrizing it. }

\medskip

 Let us call the above statement:\textit{ the quantum correlation time-symmetrization rule. }

The rule, of course, still applies to quantum algorithms where the unitary evolution builds up one-to-one quantum correlation between two initially unrelated observables (the problem-setting and the solution of the problem). In this case, it becomes redundant with the motivation for time-symmetrizing the ordinary quantum description of quantum algorithms considered until now (bringing their quantum description in line with their classical logic description). It is a redundancy that strengthens the assumption that quantum algorithms are subject to classical logic and guarantees its consequence, the need of completing their ordinary quantum description by time-symmetrizing it.

In the case instead that the building up of quantum correlation is not by a quantum algorithm, the notion of search for the solution and advanced knowledge of half of the information about the solution on the part of the problem-solver must be replaced by the more general notion of search for one-to-one correlation between the two commuting observables and reduction of the dimension of the Hilbert space in which this search takes place. This unitary building up of quantum correlation can always occur with a quadratic speedup. Indeed, the dimension of the Hilbert space in which the search for that correlation takes place still quadratically reduces, for example from $N$ to $\sqrt{N}$.

However, this time-symmetrization rule also applies to the case that the unitary evolution does not build up any quantum correlation because the two observables are maximally entangled to start with. We have seen that this is the case of quantum nonlocality.

We should eventually note that the causal loops we are dealing with should be exempt from the paradoxes that plague the usual kind. In fact, the time-symmetrization that generates them is imposed by a quantum mechanical symmetry; therefore, as long as quantum mechanics is consistent, they should be exempt from any paradox. When needed to distinguish them from the usual kind, we will call them \textit{time-symmetric causal loops} (being generated by a time-symmetrization procedure); we will call \textit{time-symmetric} also the form of retrocausality involved in them.

\section{
Discussion
}

We mention the precedents to the present work and discuss possible further developments.

\subsection{
Precedents
}
Starting with the already mentioned 1949 Wheeler and Feynman's work, various other works attempted and are attempting to go beyond the causal principle (i.e. one-way causality). However, we would think that their ways of going beyond have nothing to do with the time-symmetric causal loops of the present work. In our opinion, the discovery of the quantum speedup opens up an entirely new ground for the study of retrocausality.

There are, however, precedents to the present work, of course besides the already mentioned retrocausal approach to the quantum speedup $\left [1 -5\right ]$ of which it is the further development.

One is time-symmetric quantum mechanics $\left [17 -21\right ]$, to which it is obviously inspired. In particular, $\left [20\right ]$, about the possibility that a future choice
affects a past measurement outcome, is closely related to our final measurement of the solution changing the initial state of knowledge about the solution on the part of the problem-solver.

Another precedent concerns the incompleteness of the quantum description. We have seen that the incompleteness presently ascribed to the ordinary quantum description of quantum algorithms is exactly the same exposed by Einstein and the others in their famous EPR paper  $\left [10\right ]$. That was, of course, about the quantum description of quantum nonlocality. In $\left [5\right ]$, we showed that the causal loops generated by time-symmetrizing the ordinary quantum description of the latter exactly
play the role of the hidden variables envisaged by those authors.

A last precedent is the retrocausal explanation of quantum nonlocality given by Costa de Beauregard in 1953 $\left [11\right ]$, very similar to the present one. Consider the measurements in the same basis of the two maximally entangled observables in a state of spatial separation between the respective subsystems. Costa de Beauregard assumed that the outcome of the first measurement, performed on one of the two subsystems, locally propagates backward in time by the inverse of the unitary transformation that spatially separated the two subsystems, until these were not yet spatially separate. There, always locally, it would change the state of the other subsystem in a way that ensures the appropriate correlation between the two future measurement outcomes. The only difference with respect to our own explanation of quantum nonlocality is, of course, in the form of the retrocausality entailed. It is the retrocausality of a whole measurement outcome in Costa de Beauregard, the time-symmetric form in our case.

\subsection{
Looking forward
}
The form of retrocausality we are dealing with, which would be commonplace in ordinary quantum mechanics, would consequently have a broad impact on many aspects of science. Of special interest is that on the natural sciences, particularly on the notion of teleological evolution.

This is because the present retrocausal explanation of the quantum speedup makes of quantum algorithms full fledged teleological evolutions, indeed evolutions toward a goal (the solution of the problem) with an attractor in the very goal they will produce in the future (the solution of the problem again).

As is well known, the teleological notion has been the focus of a major debate between evolutionary biologists, physicists and natural philosophers. The question, of course, was whether the evolutions of the living can be teleological in character. Eventually, the natural sciences dismissed the notion of teleological evolution right because it implies a retrocausality deemed to be physically impossible (not so natural philosophy, see further below). That was not an easy decision. As is well known, the teleological notion perfectly matches with the functional structure of the living, for example ``teeth were created for the purpose of chewing'' (Aristotle). This difficulty is witnessed by the following sentence by the evolutionary biologist Haldane $\left [22\right ]$:

\medskip

\textit{Teleology is like a mistress to a biologist: he cannot live without her, but he is unwilling to be seen with her in public}.

\medskip

Now, the fact quantum algorithms are full fledged teleological evolutions should already shake this ``official'' position of the natural sciences. Furthermore, the fact the evolutions of the living belong to the macroscopic world described by a classical physics that cannot host any quantum speedup is not as definitive as it might seem. This is because, under the quantum cosmological assumption $\left [23 ,24\right ]$, the classical physics description of the macroscopic reality is completely avoided. This assumption is that the universe as a whole, being a perfectly isolated system since by definition there is nothing outside it, is quantum and evolves in a unitary quantum way. As a consequence, in principle, the evolution of any (also macroscopic) system can be formally obtained from that of the universe by tracing out all the rest. 

In particular, a quantum evolution of the universe would obviously involve quantum superpositions, and these, under the present theory, imply retrocausality. An immediate consequence would be the falsification of the statement that the evolution of the universe toward life cannot be teleological since this would imply a physically impossible retrocausality. Indeed, under both the quantum cosmological assumption and the present theory, the evolution of the universe toward life should host retrocausality.

Moreover, always under the quantum cosmological assumption, the teleological character of quantum algorithms becomes a strong evidence of the possibility that the teleological evolutions of life are teleological in character $\left [25\right ]$, as follows.

Indeed, for the Fine-tuned Universe version of the Anthropic Principle, the evolution of the universe toward the ``value of life'' today (either ``absence of life'' or ``existence of sentient life'') under the fundamental physical laws and their fundamental constants would acquire the form of a quantum algorithm. In fact, it would unitarily build up a (assume one-to-one) quantum correlation between the value of the fundamental constant since the beginning of time and the value of life today. This is what the quantum algorithms we have been dealing with\protect\footnote{
With the solution and invertible function of the problem-setting and determined with certainty.
} do: they unitarily build up a one-to-one correlation between the value of the problem-setting since their beginning and the value of the solution at their end. Of course, this building up can always occur with quantum speedup and consequently be teleological in character. As shown in $\left [26\right ]$, even a random quantum search for one-to-one correlation always occurs with quantum speedup. 

In the universe we are in, what could be teleological is, of course, the evolution of the universe toward life. As a consequence -- see $\left [25\right ]$ -- it could be teleological also the evolution of life on Earth (with it, the evolutions of species), the trace over all the rest of the teleological evolution of the universe toward life.

Naturally, like the quantum cosmological assumption, all the above is speculation. However, being physically plausible, it should be enough to falsify the statement that teleological natural evolutions cannot have a physical basis. Note that, if this goes against the present ``official'' position of the natural sciences on the teleological issue, it goes instead perfectly along the present position of natural philosophy, see $\left [27\right ]$. It would be a case where common sense prevailed over what, according to the present work, would turn out to have been a bias of science.

\section{Conclusions
}

The starting point of the present work is that, involving only the measurements of commuting observables -- the problem-setting and the corresponding solution -- quantum algorithms should be subject to classical logic. Therefore, their ordinary quantum description could be flanked by a classical logic description and, of course, the two descriptions should go along.

We have seen that a classical logic description of a quantum state superposition -- an essential part of quantum algorithms -- necessarily involves a logical form of retrocausality that, in turn, must be physically implicit in it.

Let us recall the reason for this. Apparently, seeing the monk on the two banks of a river at the same time is a classical logic impossibility as, logically, the two possibilities are mutually exclusive with one another. However, it is an impossibility that vanishes by resorting to retrocausality. After reaching one bank, the monk must have gone sufficiently backward in time undoing the path just followed and then forward in time again but on a different path that led him to the other bank. This identically holds in the quantum world.

Summing up, the assumption that quantum algorithms should be describable in a classical logic way brings retrocausality into ordinary quantum mechanics, where it would be as common as quantum state superpositions. The form of retrocausality we are dealing with would correspondingly be pervasive:

\medskip

It allows to provide a unified quantitative explanation of all quantum speedups and, together, quantum nonlocality.

\medskip

It shows that the building up of one-to-one quantum correlation between two initially unrelated observables that does not change the value of the observable selected by the initial measurement can always occur with quadratic quantum speedup.

\medskip

It highlights the teleological character of quantum algorithms and, under the quantum cosmological assumption, provides a plausible physical basis to the teleological character of natural evolutions.

\medskip

Starting with the famous 1949 one by Wheeler and Feynman, various attempts have been done in the past to introduce the notion of retrocausality in physics. Of course, none of them ever entered mainstream physics. We would believe that the quantum computational speedup, a relatively recent discovery, opens up an entirely new ground for the study of retrocausality. Being biased, we would think that the universal character of the retrocausality found to be implicit in the quantum computational speedup and the (apparent) evidence of its existence make it worthy of discussion.

\section*{
Acknowledgments
}
Thanks are due to Eliahu Cohen, Artur Ekert, Avshalom Elitzur and David Finkelstein for useful comments throughout the whole of the present retrocausal approach to the quantum speedup, Daniel Sheehan for organizing the San Diego AAAS Pacific Division conferences on retrocausality, a far-looking forum for the discussion of frontier, also unorthodox, physics, and Mario Rasetti for organizing the first series of international workshops on quantum communication and computation, the 1991-96  Elsag Bailey-ISI Turin Villa Gualino workshops that contributed to the dissemination of this new branch of quantum physics and led to the discovery of the major quantum algorithms.

\section*{
References
}
 $\left [1\right ]$ Castagnoli, G. and Finkelstein D. R.: Theory of the quantum speedup. \textit{Proc. Roy. Soc. Lon.}, 457, 1799-1807 (2001)

$\left [2\right ]$ Castagnoli, G.: The quantum correlation between the selection of
the problem and that of the solution sheds light on the mechanism of the quantum speed up.
\textit{Phys. Rev. A} 82, 052334 (2010)

$\left [3\right ]$ Castagnoli, G.: Completing the Physical Representation of Quantum
Algorithms Provides a Quantitative Explanation of Their Computational Speedup. Found. Phys. 48, 333-354
(2018) 

$\left [4\right ]$ Castagnoli, G.,  Cohen, E., Ekert, A. K., and  Elitzur, A. C.. A Relational
Time-Symmetric Framework for Analyzing the Quantum Computational Speedup. \textit{Found Phys.}, 49, 10 (2019),            1200-1230.

$\left [5\right ]$ Castagnoli, G.: Unobservable causal loops as a way to explain both the quantum computational speedup and quantum nonlocality. Phys. Rev. A, 104, 032203 (2021)

$\left [6\right ]$ Grover, L. K.: A fast quantum mechanical algorithm for database
search. \textit{Proc. 28th Annual ACM Symposium on the Theory
of Computing}. ACM press New York, 212-219 (1996)

$\left [7\right ]$ Wheeler, J. A. and Feynman R. P.: Classical Electrodynamics in Terms of Direct Interparticle Action. Rev. Mod. Phys. 21, 425-433 (1949)

$\left [8\right ]$ Long, G. L.: Grover algorithm with zero theoretical failure rate.
Phys. Rev. A 64, 022307-022314 (2001)

$\left [9\right ]$ Bennett, C. H.,Bernstein, E., Brassard, G., Vazirani, U.: The strengths and weaknesses of quantum computation. SIAM\ Journal on Computing, 26 (5), 1510-1523 (1997)

$\left [10\right ]$ Einstein, A., Podolsky, B., and Rosen, N.: Can Quantum-Mechanical Description of
Physical Reality Be Considered Complete? Phys. Rev. vol. 47, n. 777 (1935)

$\left [11\right ]$ Costa de Beauregard, O. La m{\'e}canique quantique. Comptes Rendus Academie des Sciences , 236, 1632-34  (1953)

$\left [12\right ]$ Finkelstein, D. R.: Space-Time Sructure in High Energy Interactions. In Fundamental Interactions at High Energy, Gudehus, T., Kaiser, and G., A. Perlmutter G. A. editors. Gordon and Breach, New York 324-338 (1969). Online PDF at http://homepages.math.uic.edu/\symbol{126}kauffman/FinkQuant.pdf

$\left [13\right ]$  Feynman, R. P.: Simulating physics with computers. Int. J. Theor. Phys. 21, 467-488 (1982)

$\left [14\right ]$ Deutsch D.: Quantum theory, the Church Turing principle and the
universal quantum computer. Proc. Roy. Soc. A 400 (1985)

$\left [15\right ]$ Rovelli, C.: Relational Quantum Mechanics. \textit{Int. J. Theor. Phys.} 35, 637-658
(1996)

$\left [16\right ]$ Fuchs, C. A.. On Participatory Realism. arXiv:1601.043603 quant-ph (2016)

$\left [17\right ]$ Aharonov, Y., Bergman, P. G., and Lebowitz, J. L.: Time Symmetry
in the Quantum Process of Measurement. \textit{Phys. Rev.} 134,
1410-1416 (1964)

$\left [18\right ]$ Dolev, S. and Elitzur, A. C.: Non-sequential behavior of the wave
function. arXiv:quant-ph/0102109 v1 (2001)

$\left [19\right ]$ Aharonov, Y. and Vaidman, L.: The Two-State Vector Formalism: An
Updated Review. \textit{Lect. Notes Phys.} 734, 399-447 (2008)

$\left [20\right ]$ Aharonov, Y., Cohen, E., and Elitzur, A. C.: Can a future choice
affect a past measurement outcome? \textit{Ann.
Phys.} 355, 258-268 (2015)

$\left [21\right ]$ Aharonov, Y., Cohen E., and Tollaksen, J.: Completely top-down
hierarchical structure in quantum mechanics. \textit{Proc. Natl. Acad. Sci. USA, }115, 11730-11735
(2018)

$\left [22\right ]$ Dronamraju, K. R.: Haldane and Modern Biology. Johns Hopkins University Press, Baltimore, Maryland  (1968)

$\left [23\right ]$ Barrow, J. D. and Tipler, F. J.: The Anthropic Cosmological Principle. 1st edition 1986 (revised 1988), Oxford University Press (1988)

$\left [24\right ]$
Bojowald, M.: Quantum cosmology. A Fundamental Description of the Universe. Lecture Notes in Physics. Vol. 835

$\left [25\right ]$
Castagnoli, G.: The physical basis of teleological evolutions. arXiv:2310.18927v13 (phys.gen-ph) (2025)

$\left [26\right ]$ Shenvi, N.,Kempe, J., Whaley, B.: A Quantum Random Walk Search Algorithm. ArXiv:quant-ph210064 (2002)

$\left [27\right ]$ Nagel, T.: Mind and Cosmos: why the materialist, neo-Darwinian conception of nature is almost certainly false. Oxford New York, Oxford University Press, ISBN 9780199919758 (2012)

\end{document}